# Integrating Atomic Scale Catalyst Design with Transport Engineering for Stable and Efficient CO$_2$ Electrolysis to CO in a Membrane Electrode Assembly

Zahra Teimouri,[1] Mahtab Masouminia,[1] Ashkan Irannezhad,[1] Reza Eslami,[1] Joseph Deering,[2] Navid Noor,[1] Shunquan Tan,[1] Amirhossein Foroozan,[1] Shayan Angizi,[1] Sung-Fu Hun,[3] and Drew Higgins [1,*]

[1]*Department of Chemical Engineering, McMaster University, Hamilton, ON, Canada*

[2] *Canadian Center for Electron Microscopy, Hamilton, ON, Canada*

[3] *Department of Applied Chemistry and Center for Emergent Functional Matter Science, National Yang Ming Chiao Tung University, Hsinchu 300, Taiwan*

**Abstract**

Electrochemical CO$_2$ reduction (CO$_2$R) offers a promising approach to decarbonize chemical manufacturing through production of carbon-neutral fuels. However, insufficient performance and instability of the membrane electrode assembly (MEA) reactors limit the commercial viability, with both metrics directly impacted by the CO$_2$R catalysts. Here we develop atomically dispersed nickel-nitrogen-carbon (NiNC) catalyst through a scalable synthesis approach using two different carbon supports. When using carbon nanotubes as a support, the resulting NiNCNT electrode achieves a partial current density towards CO of 558 mA cm$^{-2}$ with 92% Faradaic efficiency towards CO at a cell voltage of 3.2 V and an energy efficiency of 39% towards CO at a total current density of 607 mA cm$^{-2}$. The MEA demonstrated stable operation at 100 mA cm$^{-2}$ over 210 hours, outperforming previously reported NiNC catalysts. Focused ion beam-scanning electron microscopy (FIB-SEM) tomography illustrated the key role of catalyst support on the resulting performance of the electrode. COMSOL Multiphysics simulations using the 3D reconstructed images of the catalyst layers from FIB-SEM tomography demonstrated that the higher CO$_2$R performance of NiNCNT electrode is due to improved CO$_2$ diffusion and a more uniform current-density distribution compared to the NiNCB electrode prepared with carbon black as the support. The stability and performance of the NiNCNT compared favourably to the state-of-the-art Ag-based catalysts, while bottom-up cost analysis estimated the purchase cost of the NiNCNT catalyst to be $ 589 USD/kg, substantially lower than the $1,900 USD/kg estimated for Ag-based catalysts.

**Introduction**

Electrochemical $CO_2$ reduction ($CO_2R$) can store renewable electricity in the form of chemical bonds, providing a route to reduce emissions across various industrial sectors (1–3). Continuous and rapid declines in the cost of clean electricity is rendering the electricity-driven conversion technologies progressively more cost-competitive. At the same time, global energy-related $CO_2$ emissions remain very large ($\approx$ 37.8 Gt in 2024), underscoring an urgent need to reduce carbon emissions. Together, these trends position $CO_2R$ as a timely technological pathway to convert $CO_2$ into valuable chemicals and fuels, and to place $CO_2$-utilization at the center of carbon-circular manufacturing strategies (4,5). $CO_2R$ can yield a variety of products including $C_1$ (carbon monoxide, syngas ($CO/H_2$ mixture), formate, methane), and $C_{2+}$ products (ethylene, ethanol, propanol, etc.) depending on the structure and composition of the electrocatalyst, (6) as well as the local reaction environment (7). CO; a key feedstock for industrial processes like Fischer-Tropsch synthesis; is the most economically promising $CO_2R$ product due to the high selectivity and simpler separation process compared to $C_{2+}$ products (8–10). Despite considerable efforts toward the commercialization of the low-temperature (< 80°C) $CO_2R$ electrolyzers for CO production, challenges remain, including low Faradic efficiencies towards CO due to the competing hydrogen evolution reaction (HER) at high current densities (> 500 mA cm$^{-2}$) and low electrolyzer stability, which still falls short of the industrial benchmarks of up to 50,000 hours (4,11). The stability of $CO_2R$ systems is influenced by the chemical and mechanical robustness of various components, including the electrocatalysts, ionomer, and electrolyte membrane, with the electrocatalyst being recognized as a primary source of performance loss over time (12,13).

Ag and Au are considered the state-of-the-art catalysts for $CO_2R$ to CO due to their high selectivity and activity, but their high cost limits the commercial viability (14). As an alternative, catalysts consisting of atomically dispersed transition metal sites (such as Ni, Co, or Fe) coordinated by nitrogen within a carbonaceous framework (MNC) have gained significant attention due to tunable physicochemical properties, and lower costs compared to Au and Ag-based catalysts (14–16). However, a rigorous production-scale technoeconomic comparison demonstrating this cost advantage is generally lacking. Such an analysis would pinpoint dominant cost drivers such as the price of metal precursor, synthesis yield, energy/operating costs, and selection of the catalyst support, which can be translated into concrete design rules to guide scalable catalyst synthesis for CO$_2$R electrolyzers. The primary synthesis method to prepare MNC catalysts is through physical/mechanical mixing the metal and nitrogen precursors with a carbon support, followed by pyrolysis at high temperatures (>500 °C) (14,17). During the pyrolysis step, the transition metal atoms have the tendency to agglomerate to decrease their surface free energy, which enhances H$_2$ selectivity due to the propensity of metallic clusters to favour HER (18,19). On the other hand, non-uniform distribution of the nitrogen coordination sites throughout the catalyst structure can result when there is insufficient mixing of the catalyst precursors, resulting in a spatially heterogeneous distribution of the M-N$_x$ active sites that negatively affects the electrode performance (14,17,20). Despite numerous reports on the application of MNC catalysts in CO$_2$R, the impact of synthesis method, in particular the spatial distribution of the nitrogen precursors relative to the Ni precursors and carbon support, remains poorly understood.

Physicochemical properties of the carbon support (e.g., surface area, pore size distribution and hydrophobicity) used during synthesis also play a critical role in dictating the capability of MNC catalysts to meet the performance demands required for CO$_2$R scale-up: particularly, current densities (j) exceeding 200 mA cm$^{-2}$, Faradaic efficiencies (FE) towards CO exceeding 90%, and operational cell voltages less than 3 V in a MEA (21). To achieve these targets, existing literature has mainly focused on increasing the number of active sites in the catalyst or the intrinsic activity of each active site through engineering Ni-N coordination. These studies often overlook the important role of the catalyst morphology and resulting electrode structure towards the performance of CO$_2$R electrolyzers at high current density where mass transport is a critical consideration. Engineering the porosity of the catalyst layer through rational selection of the carbon support used during synthesis offers an effective approach to improve CO$_2$ transport (22). Fu et. al (23) demonstrated that NiNC electrode synthesized using biochar with hierarchical porous structure as the carbon support improved the partial current density towards CO, which originates from increased CO$_2$ transport, compared to the electrodes prepared using CB and CNT as catalyst supports in a MEA electrolyzer using vapour-fed CO$_2$ at comparable cell voltages. Beyond nanoscale porosity within the individual carbon particles, the mesoscale interparticle porosity critically controls the transport of reactants, ions and products through the catalyst layer. Well-connected pores can shorten gas-phase diffusion pathways and reduce tortuosity, increasing the effective CO$_2$ diffusivity to buried active sites to maintain high local CO$_2$ partial pressures even at high current densities. Therefore, engineering the interparticle porosity offers a direct lever to improve mass transport, selectivity and stability of the MEA electrolyzers for $CO_2R$, which is lacking in the $CO_2R$ literature (24,25).

In this work, we introduce a scalable synthesis method that controls nitrogen-precursor impregnation during synthesis to maximize the concentration of Ni-$N_x$ active sites, yielding NiNC catalyst that provide partial current density towards CO of 558 mA cm$^{-2}$ at a cell voltage of 3.2 V in a MEA reactor with a record-setting energy efficiency of 39% towards CO. We sought to understand the origins of such high performance compared to the state-of-the-art MNC and Ag catalysts by using FIB-SEM tomography to reconstruct the three-dimensional (3D) morphology of the catalyst layers prepared using either CNTs or CB with distinct morphologies as catalyst supports. This approach enables quantitative analysis of the interparticle porosity and structural connectivity in the gas diffusion electrode (GDE). The 3D-reconstructed geometries were directly incorporated into COMSOL Multiphysics simulations, enabling modeling of $CO_2$ transport based on an actual catalyst layer architecture, rather than idealized geometries. The simulations revealed that the interconnected morphology of the NiNCNT catalyst layer facilitated $CO_2$ diffusion and promoted a uniform current density distribution compared to the tortuous structure of the NiNCB catalyst layer. The production cost of the NiNCNT catalyst was also estimated as $589 USD/kg, providing an economically feasible alternative for Ag-based catalysts, which range from 1,800 to $2,000 USD/kg.

**Catalyst synthesis**

We employed two distinct routes for the synthesis of NiNC catalysts using two carbon supports: (a) carbon support (CB or CNT) was impregnated with a nickel nitrate precursor and glucose, followed by drying and physically mixing with melamine using a mortar and pestle before pyrolysis; (b) simultaneous wetness impregnation of the nickel nitrate precursor, glucose and melamine onto the carbon support, allowing chelation of the hydroxyl groups in glucose with the hydrated Ni ions (17), and π–π/H-bond interactions of melamine with the surface of the carbon support. This was followed by drying and pyrolysis (**Scheme 1**). Catalysts prepared from the first method are denoted as NiNX-phys Y, and those from the second methodology as NiNX-sim Y, where X stands for CNT or CB and Y represents the pyrolysis temperature.

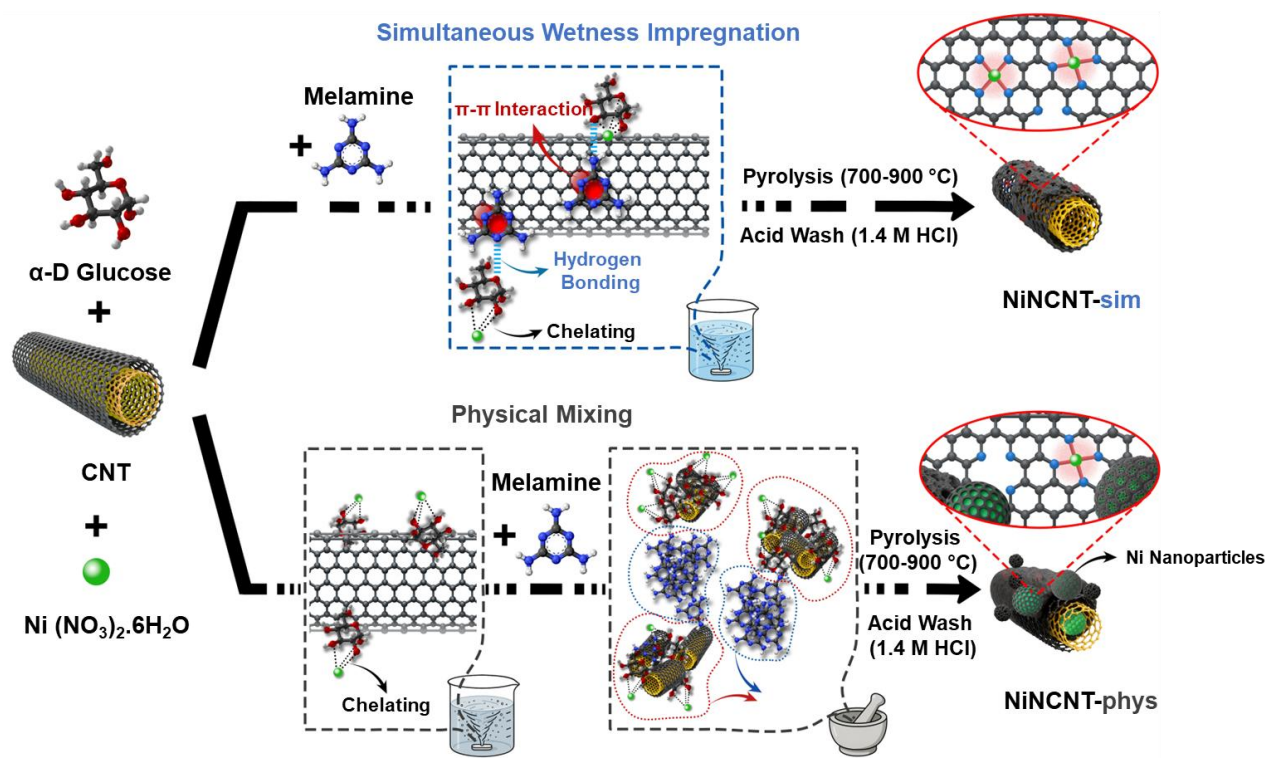

**Scheme 1.** Schematic illustration of the synthesis process for simultaneous wetness impregnation and physical mixing of melamine for synthesis of the NiNCNT catalysts.

**CO$_2$R performance results**

The CO$_2$R performance of the NiNC electrodes was investigated in a MEA, with additional details on the experimental set-up provided in the **Methods** section. We investigated the impact of pyrolysis temperature on the CO$_2$R performance of the synthesized electrodes (**Fig. 1a-c**). The NiNCNT-sim 800 °C electrode (**Fig. 1b**) exhibits higher j$_{CO}$ than the NiNCNT-sim 700 °C (**Fig. 1a**) and NiNCNT-sim 900 °C electrodes (**Fig. 1c**) at different cell voltages that were tested, reaching to a j$_{CO}$ of 558 mA cm$^{-2}$ at 3.2 V with a FE$_{CO}$ of 92%. We also studied how catalyst synthesis method controls the CO$_2$R performance of the electrode in a MEA. The NiNCNT-phys electrodes (**Fig. 1d-f**) provided a lower j$_{total}$ and FE$_{CO}$ compared to the catalysts prepared using simultaneous wetness impregnation method (NiNCNT-sim electrodes) at all tested cell voltages (**Fig. 1a-c** and **Fig. 2a**).

The operational stability of the best performing electrode (NiNCNT-sim 800 °C) was evaluated using chronopotentiometry measurements at a constant current density of 100 mA cm$^{-2}$, with the results shown in **Fig. 2b**. The CO$_2$R performance was largely retained over 210 hours of electrolysis. Particularly, less than a 5% change in the cell voltage and FE towards CO was observed throughout the operation. This stability is over three times longer than the maximum of 70 hours previously reported in the literature for state-of-the-art NiNC catalysts tested in a MEA (21,26–28). Compared to previous NiNC electrodes tested in a MEA with current densities > 500 mA cm$^{-2}$, the NiNCNT-sim 800 °C electrode provides the highest energy efficiency reported to date towards CO production, reaching ~ 39% (**Table S1** and **Fig. 2c**). We also compared the CO$_2$R performance and stability of the NiNCNT-sim 800 °C electrode with state-of-the-art Ag electrodes operating in a MEA (**Fig. 2d** and **Table S1**) (29–34). Compared with Ag electrodes operating between 3.2–3.6 V, our best performing electrode (NiNCNT-sim 800 °C) achieved ~15–20% higher j$_{CO}$ at equal or lower cell voltage with higher energy efficiency towards CO. Although one Ag electrode reaches to a higher j$_{CO}$ of 700 mA cm$^{-2}$, it does so with a lower FE$_{CO}$ (85%) and EE$_{CO}$ (32.5%). Taken together, our data establish a best-in-class reaction rate-efficiency-stability balance, pairing a high current density with a low cell voltage, exceeding 90% FE$_{CO}$ and EE$_{CO}$ of 39%.

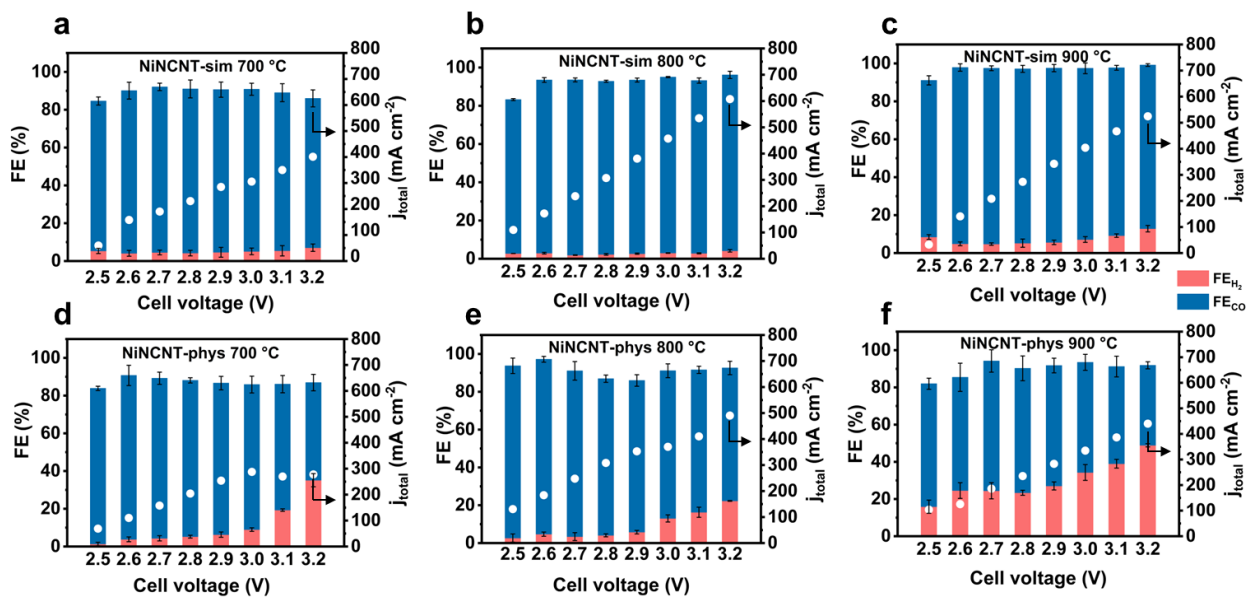

**Fig. 1** | CO$_2$R performances of: **a,** NiNCNT-sim 700 °C, **b,** NiNCNT-sim 800 °C, **c,** NiNCNT-sim 900 °C, **d,** NiNCNT-phys 700 °C, **e,** NiNCNT-phys 800 °C, **f,** NiNCNT-phys 900 °C eelctrodes. The CO$_2$R performance tests were conducted in a MEA using 0.5M KHCO$_3$ as anolyte. The error bars in **a-f**, correspond to the standard deviation of three independent measurements.

One key reason for 210 hours of stability for the NiNCNT-sim 800 °C electrode (**Fig. 2b**) could potentially be due to the incorporation of polytetrafluoroethylene (PTFE) particles into the electrode structure, which has been shown to mitigate flooding of the electrode by modifying the hydrophobicity (35). Flooding refers to the blockage of the GDL pores with liquid, hampering $CO_2$ transport (36). Static contact angle measurements revealed a highly hydrophobic surface (144.35°) for the NiNCNT-sim 800 °C electrode (**Fig. S1b**). The stability test for NiNCNT-sim 800 °C electrode eventually led to 16% FE towards $H_2$ after 230 hours of chronopotentiometry at 100 mA cm$^{-2}$. We then sought to further improve the stability of this electrode by operating the cell under pulsed conditions through alternating between chronopotentiometry at the target current density for 3s followed by 0 mA cm$^{-2}$ for 1s. When utilizing a current density of 200 mA cm$^{-2}$, selected to study the stability of the electrode at more industrially relevant conditions, the NiNCNT-sim 800 °C electrode maintained the performance for 70 hours, showing only a 10% change in the cell voltage and a 11% decrease in FE towards CO. When utilizing a current density of 100 mA cm$^{-2}$ under pulsed conditions, the MEA sustained continuous operation for 300 hours with just 11% and 17% changes in the cell voltage and $FE_{CO}$, respectively (**Fig. S2a, b**). Improved stability of the NiNCNT-sim 800 °C electrode under pulsed conditions compared to the continuous galvanostatic operation at 100 mA cm$^{-2}$ as demonstrated in **Fig. 2b** (210 hours of stability) and **Fig. S2b** (300 hours of stability), can likely be associated to the regulation of catalyst microenvironment during the recovery periods of the pulsed technique. In the recovery periods (an applied current density of 0 mA cm$^{-2}$ for 1 s), anions such as $CO_3^{2-}$ formed because of the reaction between $OH^-$ ions (produced in $CO_2R$ and HER) with $CO_2$ gas, can diffuse away from the catalyst surface. This periodic adjustment helps to suppress salt precipitation, mitigating performance degradation (37,38).

Although pulsed electrolysis utilizing a current density of 200 mA cm$^{-2}$ enabled relatively stable performance of the NiNCNT-sim 800 °C electrode in a MEA over 70 hours; degradation ultimately emerged after ~70 hours as the FE towards $H_2$ increased to 12% with cell voltage of 3.05 V compared to the starting point with $FE_{H2}$ of only 1% and cell voltage of 2.75 V (**Fig. S2a**). We then explored the possible contributions of the reversible performance losses that may originate from salt precipitation and/or flooding. After 70 hours of $CO_2R$, we rinsed the NiNCNT-sim 800 °C electrode to remove any possible salts that had formed, followed by drying the electrode to reverse the impacts of flooding. After re-assembly and applying chronopotentiometry testing at 200 mA

cm$^{-2}$, the FE towards H$_2$ decreased to 8.2% (**Fig. S3**) from the 12% recorded after 70 hours of stability test, which enabled us to arrive at this conclusion that a substantial fraction of the degradation is reversible and attributable to the system-level failure modes such as salt deposition and flooding rather than the irreversible catalyst deactivation.

The CO$_2$R performance of the CNT supported catalyst prepared via the simultaneous wetness impregnation method without the addition of nickel nitrate (NCNT-sim 800 °C) was also investigated and found to provide a maximum j$_{CO}$ of only 20 mA cm$^{-2}$ at a cell voltage of 2.7 V (**Fig. S4a**), highlighting the key role of Ni in creating catalytically active CO$_2$R sites, that are likely in the form of Ni-N$_x$ as shown by previous literature (20). It is postulated that the production of small amounts of CO in the Ni-free NCNT-sim 800 °C electrode likely arises from the residual transition metals (e.g., Co, Fe) present in the structure of the CNT or trace metal impurities in the precursors used during catalyst synthesis, which upon pyrolysis can potentially form CoNC and FeNC active sites that are known to be catalytically active, an occurrence that was previously brought to light within the field of oxygen reduction electrocatalysis (39,40). We also evaluated the impact of using glucose as a chelating agent during catalyst synthesis and observed a clear performance gain: at 3.2 V, the NiNCNT-sim 800 °C electrode delivered a j$_{CO}$ of 558 mA cm$^{-2}$ with FE$_{CO}$ of 92%, outperforming the electrode prepared without glucose (NiNCNT-sim 800 °C w/o glucose) with a j$_{CO}$ of 402 mA cm$^{-2}$ and FE$_{CO}$ of 89% (**Fig. S4b, c**).

We also probed the impact of catalyst synthesis on the CO$_2$R performance of the electrodes prepared using CB as the catalyst support. The NiNCB-sim 800 °C electrode demonstrated a j$_{CO}$ of 154 mA cm$^{-2}$ at 3.2 V and FE$_{CO}$ of 75%, a significant improvement over the NiNCB-phys 800 °C electrode with a j$_{CO}$ of 83 mA cm$^{-2}$ and FE$_{CO}$ of 55% at 3.2 V (**Fig. S5a, b**). As the porosity and CO$_2$ transport within the electrode can be tuned by engineering the catalyst support, we compared the CO$_2$R performance of the NiNCNT and NiNCB electrodes within a MEA. The NiNCNT electrodes provided significantly higher j$_{total}$ and j$_{CO}$ values across a broad cell voltage window (2.5-3.2 V) compared to the NiNCB electrodes (**Fig. 1** and **Fig. S5a-c**). For example, the NiNCNT-sim 800 °C electrode demonstrated a j$_{CO}$ of 558 mA cm$^{-2}$ and FE$_{CO}$ 92% at 3.2 V, outperforming the NiNCB-sim 800 °C electrode with a j$_{CO}$ of 154 mA cm$^{-2}$ and FE$_{CO}$ of 75% at the same operating cell voltage as illustrated in **Fig. 1b** and **Fig. S5a**, respectively. In line with the NiNCNT electrodes, the NiNCB electrodes display the same pyrolysis-temperature-dependent performance, where NiNCB-sim 800 °C (with $j_{CO}$ of 154 mA cm$^{-2}$ at 3.2 V) surpasses NiNCB-sim 900 °C ($j_{CO}$ of 99 mA cm$^{-2}$ at 3.2 V) (**Fig. S5a, c**).

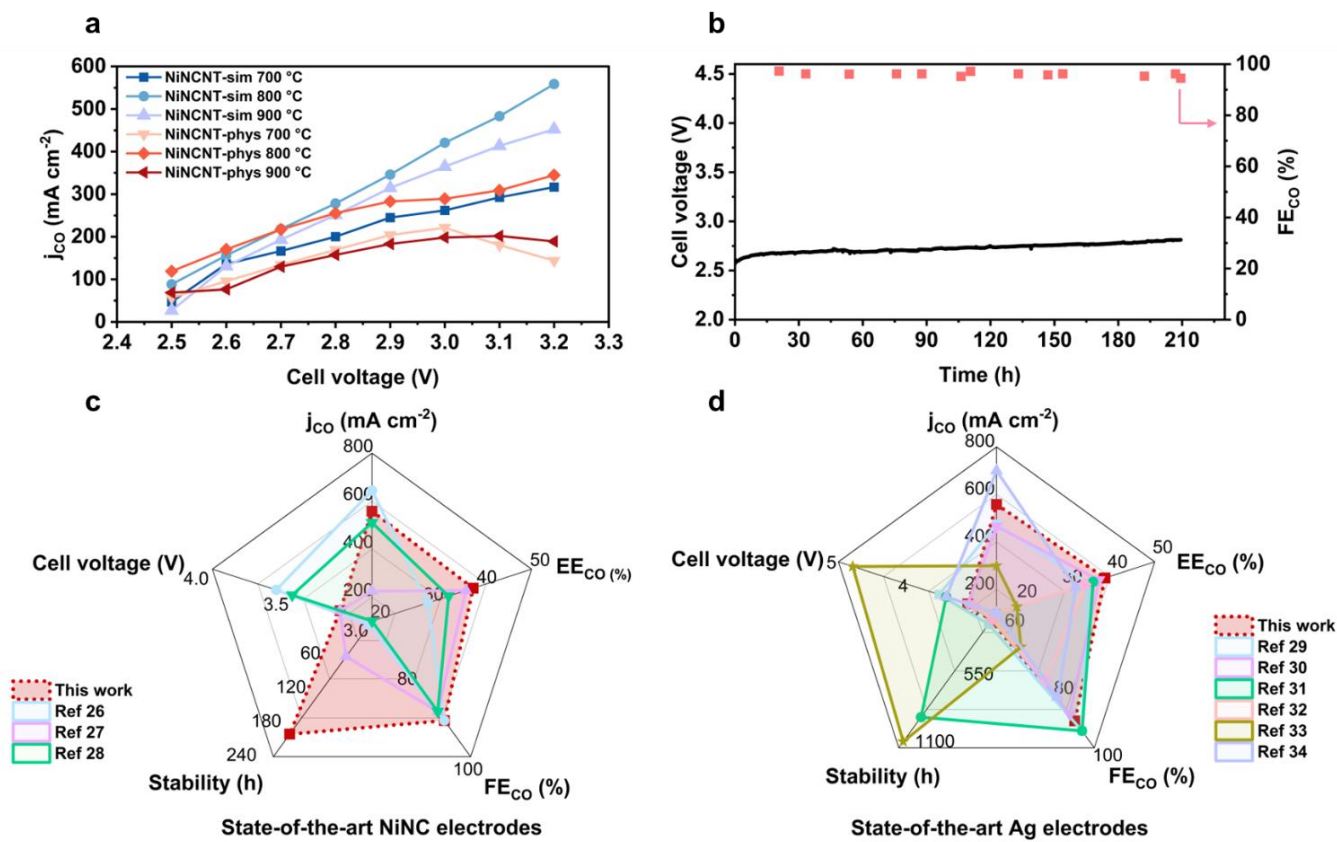

**Fig. 2** | **a,** Effect of pyrolysis temperature and catalyst preparation method on the partial current density towards CO, **b,** stability of the NiNCNT-sim 800 °C electrode in a MEA at applied current density of 100 mA cm$^{-2}$ using 0.1M KHCO$_3$ electrolyte, Comparison of the EE$_{CO}$, FE$_{CO}$, cell voltage, stability and $j_{CO}$ in this work with those of the state-of-the-art **c,** NiNC electrodes (26–28), and **d,** Ag electrodes in a MEA (29–34).

**Catalyst Characterization**

The Brunauer-Emmett-Teller (BET) surface area and pore volume were evaluated for acid-treated CNT, acid-treated CB, NiNCNT-sim, NiNCNT-phys, NiNCB-sim, and NiNCB-phys catalysts synthesized at pyrolysis temperatures ranging from 700 to 900 °C (**Fig. 6a-f** and **Table S2**). The N$_2$ adsorption-desorption isotherm of the acid-treated CNT shown in **Fig. S6f** corresponds to a type IV isotherm, while acid-treated CB exhibits a combined type I + IV isotherm, indicating the presence of both microporous and mesoporous structures (41). The NiNCNT-sim 800 °C catalyst demonstrated higher specific surface area (596 m$^2$ g$^{-1}$) compared to the acid-treated CNT (486 m$^2$

g$^{-1}$). This can be correlated to the presence of Ni during catalyst synthesis, which likely promotes microporosity development by catalyzing carbonization of the glucose/melamine during pyrolysis (42). By increasing the pyrolysis temperature to 900 °C; the specific surface area slightly decreased for the NiNCNT-sim 900 °C catalyst (422 m$^2$ g$^{-1}$) compared to the NiNCNT-sim 800 °C catalyst (596 m$^2$ g$^{-1}$) as shown in **Table S2**. On the other hand, the catalyst prepared using physical mixing method (NiNCNT-phys 800 °C) demonstrated a lower specific surface area (373 m$^2$ g$^{-1}$) compared to the acid-treated CNT (486 m$^2$ g$^{-1}$). This suggests that the inhomogeneous dispersion of melamine may have partially blocked inter-CNT spaces, reducing the specific surface area.

To probe how the carbon support and synthesis route impact the physical properties of the NiNC catalyst, we also analyzed the BET surface areas of the NiNCB catalysts. NiNCB-sim 800 °C showed a markedly higher specific surface area (1,019 m$^2$ g$^{-1}$) than both the acid-treated CB (758 m$^2$ g$^{-1}$) and NiNCB-phys 800 °C catalyst (645 m m$^2$ g$^{-1}$), highlighting that the simultaneous wetness impregnation method promotes additional porosity development in the catalyst structure during pyrolysis. The enhanced surface area of the NiNCB-sim 800 °C and NiNCNT-sim 800 °C catalysts (**Table S2**) compared to the NiNCB-phys 800 °C, and NiNCNT-phys 800 °C counterparts, correlate well with their higher CO$_2$R activity and selectivity (**Fig. S5a, b** and **Fig. 1b, e**). Higher specific surface area can expose more electrochemically active sites and enhance the porosity of catalyst that can facilitate CO$_2$ transport. The impact of glucose on the physical characteristics of the NiNC catalyst was also supported by BET data. Improved CO$_2$R performance observed for the NiNCNT-sim 800 °C electrode (**Fig. 1b**) compared to the NiNCNT-sim 800 °C w/o glucose electrode (**Fig. S4b**) can also be attributed to the formation of abundant micropores (<2 nm) originated from glucose carbonization during pyrolysis, as evidenced by pore-size distribution analysis and micropore surface area in the NiNCNT-sim 800 °C catalyst (**Fig. S6b** and **Table S2**).

High-resolution TEM (HRTEM) images (**Fig. 3a, b**) reveal that the NiNCNT-sim 800 °C structure is covered with a partially amorphous and partially crystalline surface layers, which is absent in the structure of the pristine CNT as shown in **Fig. 3c**. This overlayer is a typical feature of the pyrolyzed carbonaceous materials such as glucose (17). To further evaluate the characteristics of this overlayer; we compared the diffraction patterns of this layer in the NiNC catalysts with the pristine supports (CB and CNT). Fast Fourier transform (FFT) patterns for the NiNCNT-sim 800

°C (**Fig. 3a, b**) and NiNCB-sim 800 °C (**Fig. S7a**) show diffuse rings, supporting the presence of an isotropic carbon overlayer derived from partial carbonization of glucose/melamine during pyrolysis, whereas acid-treated CNT (**Fig. 3c**) and CB (**Fig. S7b**) demonstrate arc-like orientation, which is characteristics of the well-ordered sp$^2$ carbon (002). Z-contrast high-angle annular dark-field and aberration-corrected scanning transmission electron microscopy (HAADF-STEM) images exhibited isolated bright spots on the CNT (**Fig. 3d-f**), which are suggested to be Ni based on the Ni L-edge signal obtained by the electron energy loss spectroscopy (EELS, **Fig. 3g, h**). The presumably atomically dispersed Ni sites are also anchored on the carbon layer derived from the glucose/melamine carbonization as wells as on the sidewalls of CNT (**Fig. 3d**). STEM-EDX (energy dispersive X-ray spectroscopy) elemental mapping further shows the distribution of Ni and N throughout the carbon support for NiNCNT-sim 800 °C (**Fig. 3i**) and NiNCB-sim 800°C (**Fig. S7e**) catalysts.

To further analyze the impact of synthesis method, chelating agent, and carbon support on the structure of the synthesized NiNC catalysts; we used Raman spectroscopy. Raman spectra show the D and G band peaks for all catalysts located at ~1350 cm$^{-1}$ and 1590 cm$^{-1}$, respectively (**Fig. S8a, b**). NiNCNT-sim 800 °C catalyst shows a lower $I_D/I_G$ ratio and a clear 2D band at ~2700 cm$^{-1}$ compared to the NiNCNT-phys 800 °C catalyst. We observed the same trend for the catalysts prepared using CB support as NiNCB-sim 800 °C catalyst demonstrated a lower $I_D/I_G$ ratio compared to NiNCB-phys 800 °C catalyst (**Fig. S8b**). NiNCNT-sim 800 °C w/o glucose shows a higher $I_D/I_G$ indicating that glucose promotes graphitization of the carbon; FFT patterns further corroborate the partially graphitized nature of the glucose/melamine-derived carbon overlayer as shown in **Fig. 3b**. The X-ray diffraction (XRD) also displays two broad peaks associated to the (002) and (101) planes of the graphitic carbon with no metallic Ni peaks observed for any of the NiNC catalysts, likely due to the acid washing step implemented during catalyst synthesis (**Fig. S8c**).

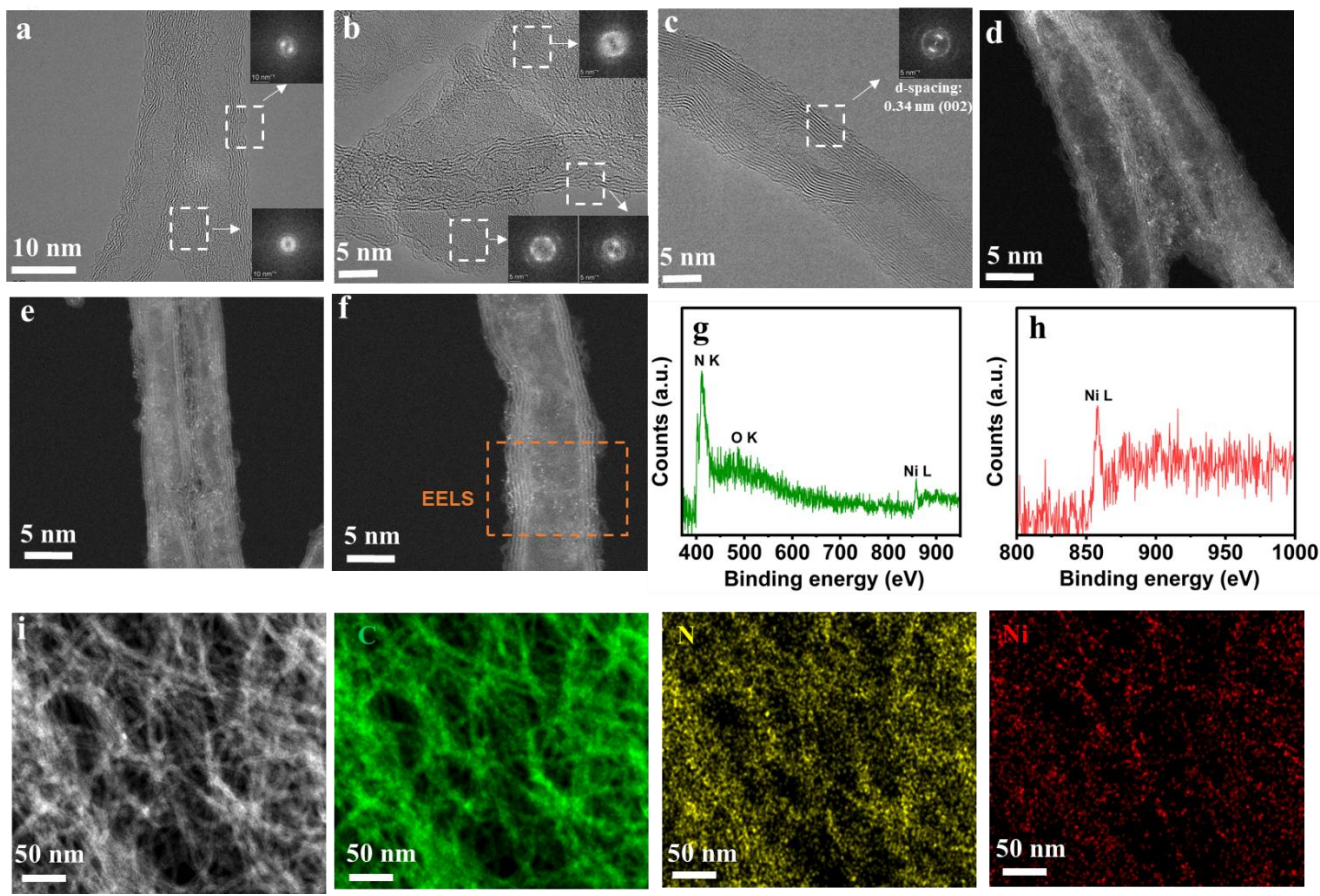

**Fig. 3 |** Morphological and compositional analysis of NiNCNT-sim 800 °C catalyst. **a, b,** HRTEM image of NiNCNT-sim 800 °C **c,** HRTEM image of acid-treated CNT **d-f,** HAADF-STEM images of NiNCNT-sim 800 °C catalyst **g, h,** EELS spectra of NiNCNT-sim 800 °C catalyst, and **i,** EDX map of NiNCNT-sim 800 °C catalyst.

X-ray photoelectron spectroscopy (XPS) was also performed to assess the effects of synthesis method, carbon support, and pyrolysis temperature on the concentration and identity of the resulting N functionalities. Elemental quantification results derived from XPS shown in **Table S3** reveal higher Ni atomic percentages for NiNCNT-sim 800 °C catalyst (0.53 %) compared to NiNCNT-phys 800 °C (0.42 %). The high-resolution N 1s spectra (**Fig. 4a-d**) was deconvoluted into pyridinic N (~398.5 eV), Ni-N (~399.2 eV), pyrrolic N (~400.2 eV), graphitic N (~401.8 eV), and oxidized N (~406.7 eV) (23). As depicted in **Fig. 4e**, both the proportion of N atoms residing in pyridinic form and the total atomic percentage of N in the catalyst were reduced by applying increased pyrolysis temperature ranging from 700 to 900 °C for NiNCNT-sim catalysts.

To further elucidate the role of glucose during catalyst synthesis in terms of maintaining Ni in atomically dispersed form to produce Ni-$N_x$ sites, the N1s peak deconvolution of the NiNCNT-sim 800 °C w/o glucose was also performed. NiNCNT-sim 800 °C w/o glucose showed a lower proportion of N atoms residing in the Ni-N form (10.54%), as well as a lower total atomic percentage of N (0.96%) compared to the NiNCNT-sim 800 °C catalyst with total atomic percentage of N of 1.21% and 24.97% proportion of N atoms residing in the Ni-N form as shown in **Table S3** and **Fig. 4d**. Both catalysts were synthesized with the same nominal Ni loading (0.69 wt% of the total precursor mixture including CNT, glucose and melamine). This suggests the formation of small Ni clusters in the absence of glucose during pyrolysis. Any Ni clusters that were formed probably have been leached out during the acid washing step (**Scheme 1**), thereby resulting in the reduced final Ni concentration in the glucose-free sample. Consistent with previous reports (43,44), lower N contents in the catalysts with the increase of pyrolysis temperature does not correspond to a decline in $CO_2R$ performance of the NiNC catalysts (**Fig. 1**). On the other hand, the proportion of N atoms residing in the Ni-N form illustrates an upward trend by increasing pyrolysis temperature. This suggests that the catalytic activity in $CO_2R$ is governed primarily by the specific chemical nature and local environment of the nitrogen functional groups, rather than their overall abundance.

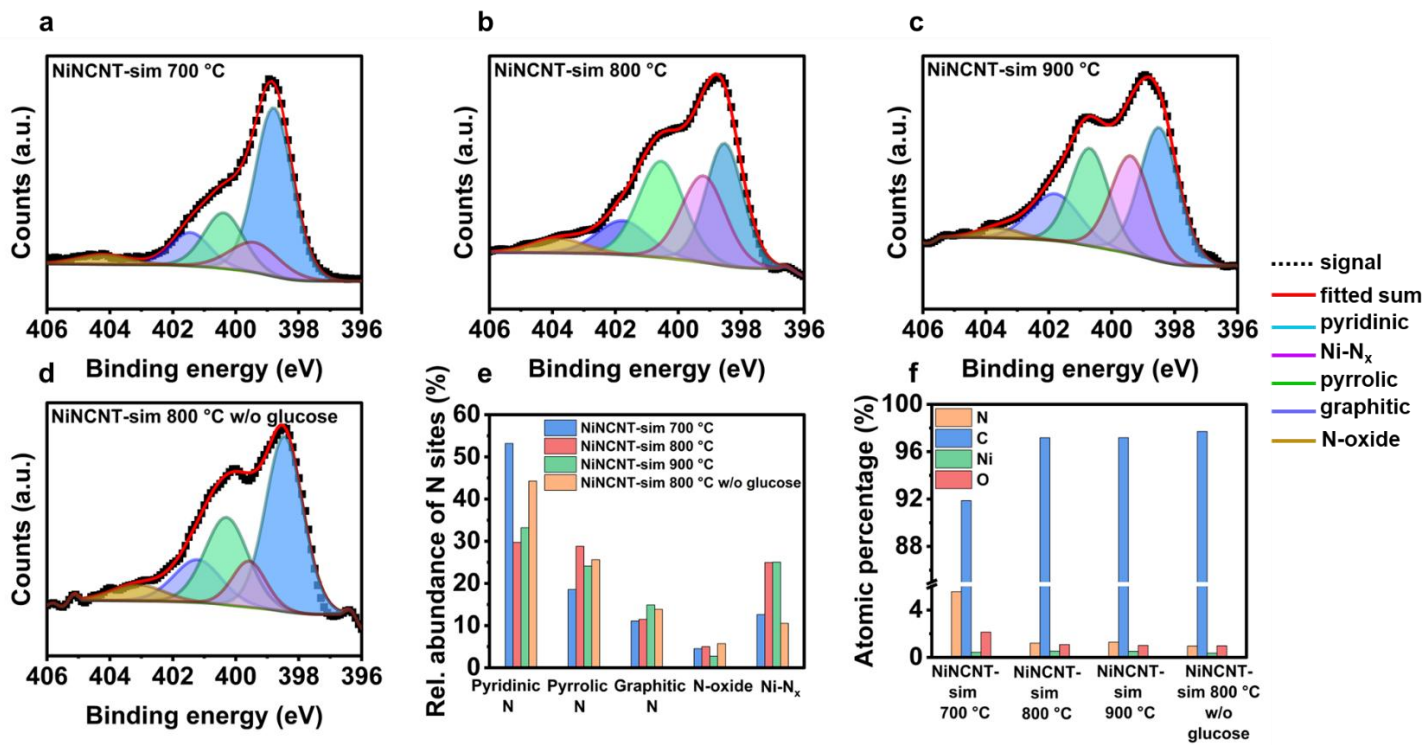

**Fig. 4** | Deconvoluted high resolution N 1s XPS spectra of **a,** NiNCNT-sim 700 °C, **b,** NiNCNT-sim 800 °C **c,** NiNCNT-sim 900 °C, **d,** NiNCNT-sim 800 °C w/o glucose, **e,** relative abundance of different N sites in the NiNCNT catalysts, and **f,** atomic percentages of N, C, Ni, and O in the synthesized NiNCNT-sim catalysts.

The local coordination environment of Ni atoms in the prepared catalysts was analyzed using X-ray absorption near-edge structure (XANES) and extended X-ray absorption fine structure (EXAFS) analyses. The XANES spectrum of NiPc as a reference demonstrates a weak pre-edge feature at position A and a strong absorption peak at position B (**Fig. 5a**). The former arises from a local electric quadrupole transition (1s → 3d), while the latter corresponds to the dipole-allowed transition (1s → 4$p_z$) characteristic of Ni-$N_4$ centers (45). The main edge position of the NiNCNT-sim, NiNCNT-phys, NiNCB-sim, and NiNCB-phys catalysts synthesized at different pyrolysis temperatures is slightly higher than the Ni foil and lower than NiO (**Fig. 5a, b** and **Fig. S9**); highlighting that the oxidation state of Ni is between 0 and +2, which is characteristic of atomically dispersed NiNC catalysts (19).

Fourier transform (FT) $k^2$-weighted EXAFS spectra of the synthesized catalysts were studied in R space (**Fig. 5c-f**). As shown in **Fig. 5e** (a magnified view of **Fig. 5c**), the dominant peak in the Ni K-edge EXAFS-FT spectra of the NiNCNT-sim 700 °C and NiNCNT-sim 800 °C catalysts appear at approximately 1.45 Å (non-phase-shift-corrected), which is attributed to the Ni-N scattering path within Ni-N$_4$ coordination. The major peak in the EXAFS-FT spectra of the NiNCNT-sim 900 °C on the other hand, was observed at 1.38 Å, highlighting that the average Ni-N radial distance might be shorter than that of the NiPc (~1.45 Å), demonstrating different Ni-N coordination in the NiNCNT-sim 900 °C compared to NiPc, and the NiNCNT catalysts pyrolyzed at 700 and 800 °C. This was further supported by EXAFS fitting, which demonstrates that the Ni-N coordination in the NiNCNT-sim 700 °C, NiNCNT-sim 800 °C, and NiNCNT-sim 900 °C are 4.08, 4.12, and 3.32, respectively (**Table S4** and **Fig. S10**). A similar trend was also reported by Wang et al. (20), showing that increasing pyrolysis temperature from 800 to 900 °C for MNC catalysts (M=Fe, Ni, Co, Cu, and Mn) led to a shorter M-N radial distance. In addition to the primary Ni-N scattering peak, the EXAFS-FT spectrum of the NiNCNT-sim 800 °C w/o glucose catalyst exhibits a shoulder near ~2.2 Å (**Fig. 5d**), likely arising from Ni-Ni scattering associated with the presence of Ni clusters in the catalyst synthesized without glucose. Notably, such short-range Ni-Ni order can be invisible by XRD as shown in **Fig. S8c**, highlighting the key role of EXAFS in evaluating the local structure of the synthesized catalysts. This is also in line with the HER activity of the NiNCNT-sim 800 °C w/o glucose electrode (**Fig. S4b**), which can be correlated to the presence of small Ni clusters (46).

We also analyzed the local chemical structure of the catalysts prepared using CB support. NiNCB-sim 800 °C, NiNCB-sim 900 °C, and NiNCB-phys 800 °C catalysts display a main peak at about 1.41 Å corresponding to the first-shell Ni-N coordination (**Fig. 5f**). The similar peak position to that of the NiPc suggests that the Ni centers in NiNCB catalysts are primarily coordinated in the form of Ni-N$_4$ geometry (**Table S4**). For NiNCB-phys 800 °C catalyst, an additional peak at 2.2 Å was also found in the FT-EXAFS spectrum (**Fig. 5f**) depicting the existence of small Ni clusters (20). This likely arises from the non-uniform distribution of melamine, which limits the availability of N coordination sites and promotes Ni clustering. This is also consistent with the CO$_2$R performance results showing a higher HER selectivity for NiNCB-phys 800 °C catalyst compared to NiNCB-sim 800 °C catalyst (**Fig. S5a, b**). The wavelet transformed (WT) plots of the reference samples (NiO, NiPc, and Ni foil), and NiNCNT-sim 800 °C catalyst is also presented in **Fig. 5g-k**. NiNCNT-sim 800 °C catalyst shows a maximum lobe at 5.8 Å$^{-1}$, assigning to the Ni-N bonding when compared to NiPc. No intensity maximum of Ni-Ni bonding is detected for NiNCNT-sim 800 °C catalyst at higher wave numbers.

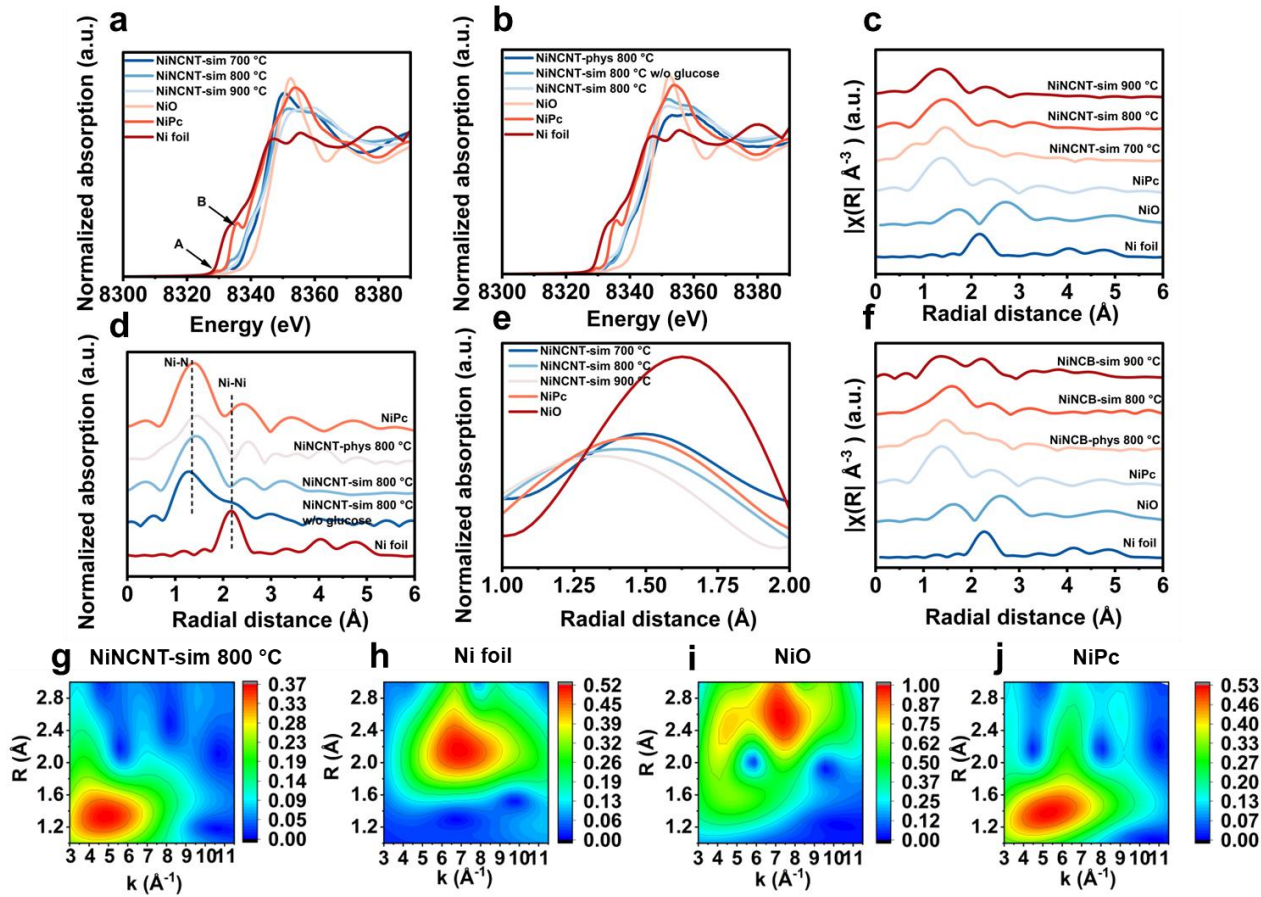

**Fig. 5 |** Chemical states and atomic coordination environment analysis of synthesized catalysts. XANES spectra of the **a, b,** NiNCNT catalysts synthesized at different pyrolysis temperatures, melamine impregnation method, and without glucose, **c-f,** FT-EXAFS spectra at the Ni K-edge of NiNC catalysts synthesized at different pyrolysis temperatures, melamine impregnation method, without glucose, and different carbon substrates, Wavelet transforms (WT) of **g,** NiNCNT-sim 800 °C, **h,** Ni foil, **i,** NiO, and **j,** NiPc.

*In situ* electrochemical impedance spectroscopy (EIS) was employed to study the charge transfer, electrolyte resistance, and mass transport in the prepared electrodes. Distribution of the relaxation times (DRT) analysis was conducted by deconvoluting the impedance data to correlate the electrochemical reactions with their corresponding relaxation times (47). As the applied cell voltage increases, the charge transfer resistance ($R_{ct}$) decreases for the NiNCNT-sim 800 °C (**Fig. S11a**), NiNCNT-sim 800 °C w/o glucose (**Fig. S11b**), NiNCNT-phys 800 °C (**Fig. S11c**), NiNCB-phys 800 °C (**Fig. S12a**), and NiNCB-sim 800 °C (**Fig. S12b**) electrodes. The correlated DRT plots show 3 distinguished peaks as demonstrated in **Fig. S11d-f**. The $P_1$ peak located at approximately $2.0 \times 10^{-6}$ s shows an inconspicuous variation with cell voltage, suggesting the polarization represented by this peak is irrelevant to the $R_{ct}$ of $CO_2R$ and it is more correlated to contact resistance. The $P_2$ peak (~$2.0×10^{-5}$ s) on the other hand, gradually diminishes with increasing cell voltage (**Fig. S11d**), whereas $j_{CO}$ rises sharply by increasing the cell voltage from 2.5 to 3.2 V (**Fig. 1b**). This opposing trend suggests $P_2$ is unlikely to reflect the $R_{ct}$ of $CO_2R$ and is more consistent with a process associated with the HER. The peak at about $1.0 × 10^{-3}$ s ($P_3$) can be attributed to the $CO_2$ charge transfer in the NiNCNT-sim 800 °C electrode (**Fig. S11d**).

To further evaluate the impact of glucose; we studied the electrochemical behaviour of NiNCNT-sim 800 °C w/o glucose electrode. At 3.2 V, selected as a reference condition where mass transfer limitations are more likely to arise, the resistance value of $P_3$ in the NiNCNT-sim 800 °C w/o glucose electrode is slightly higher than the NiNCNT-sim 800 °C electrode, suggesting higher charge transfer resistance for the electrode prepared without glucose (**Fig. S11e**). We also investigated the role of catalyst synthesis method on the electrochemical behaviour of the synthesized NiNC electrodes. NiNCNT-phys 800 °C electrode indicated higher charge transfer resistances compared to the NiNCNT-sim 800 °C electrode at different cell voltages (**Fig. S11a, c**), which are also in line with the DRT results. $P_3$ peak shifted to lower relaxation times and amplitude of the impedance, indicating improved $CO_2R$ kinetics for NiNCNT-sim 800 °C compared to the NiNCNT-phys 800 °C electrode (**Fig. S11f**). Similarly, NiNCB-sim 800 °C electrode depicted moderately lower $R_{ct}$ at higher cell voltages compared to NiNCB-phys 800 °C electrode as demonstrated in **Fig. S12a, b**. These highlight improved interfacial kinetics in the NiNCB-sim 800 °C electrode, which is correlated with the better electronic coupling between Ni-$N_x$ sites and the carbon backbone, fewer resistive contact points within the catalyst layer, and/or a more favorable local reaction environment (wettability/ion transport) at the most active domains. Possible inhomogeneous dispersion of the Ni-$N_x$ sites derived from the physical mixing of the N precursor in the catalyst synthesis could also potentially increase the interfacial charge transfer resistance. We then sought to investigate the impact of catalyst support on the interfacial charge transfer/charge transfer resistance of the NiNC electrodes. Based on the DRT results at 3.2 V as illustrated in **Fig. S12c**, the polarization resistance value of $P_3$ decreased for NiNCNT-sim 800 °C electrode compared to NiNCB-sim 800 °C, indicating improved charge transfer for NiNCNT-sim 800 °C electrode.

To analyze the effective interfacial area of the catalyst layer wetted by electrolyte, we used cyclic voltammetry (CV) measurements to calculate the double-layer capacitance ($C_{dl}$). The NiNCNT-sim 800 °C catalyst indicated higher $C_{dl}$ (6.59 mF cm$^{-2}$) compared to NiNCNT-sim 800 °C w/o glucose (4.40 mF cm$^{-2}$) catalyst as illustrated in **Fig. S13a, b** and **Fig. S14,** respectively. We noted two chief reasons why the NiNCNT-sim 800 °C catalyst shows higher double layer capacitance compared to the NiNCNT-sim 800 °C w/o glucose catalyst: (1) formation of the carbon layer derived from glucose carbonization (supported by BET results (**Table S2**) and HRTEM images (**Fig. 3b**)), which creates additional micropores and increases the electrochemically exposed surface area; and (2) impact of the glucose in isolating Ni sites during pyrolysis through chelation of the hydrated Ni ions with hydroxyl groups of the α-D-glucose molecule as visualized in **Scheme 1**. A similar chelation effect has been previously reported for FeNC atomically dispersed catalyst for oxygen reduction reaction (ORR) (17). Notably NiNCNT-phys 800 °C catalyst showed lower $C_{dl}$ (3.40 mF cm$^{-2}$) compared to the NiNCNT-sim 800 °C catalyst (6.59 mF cm$^{-2}$), underscoring the beneficial effect of simultaneous wetness impregnation of melamine during synthesis on the electrolyte-electrode interactions. This trend is consistent with the lower BET surface area of the NiNCNT-phys 800 °C (373 m$^2$ g$^{-1}$) relative to the NiNCNT-sim 800 °C (596 m$^2$ g$^{-1}$) catalyst, suggesting restricted pore accessibility arising from non-uniform melamine dispersion during physically mixed method, which leads to Ni agglomerates remaining in the structure of the catalyst even after acid washing (**Scheme 1**).

**Tomographic analysis**

To evaluate the impact of catalyst support and interparticle porosity on the $CO_2R$ performance of the NiNCNT-sim 800 °C and NiNCB-sim 800 °C electrodes in MEA; we employed FIB-SEM tomography imaging to identify the morphology of the catalyst layer, which was then coupled with COMSOL Multiphysics to evaluate the morphological impacts on the $CO_2$ transport (**Fig. 6a**, **Supplementary Note 1**, and **Fig. S15**). 3D reconstructions from FIB-SEM tomography (**Fig. 6b, c**) show that the NiNCNT-sim 800 °C catalyst particles are anisotropic and fibrous, whereas the NiNCB-sim 800 °C catalyst particles are isotropic and tortuous. Quantitative pore analysis demonstrates that the NiNCNT-sim 800 °C catalyst layer attains a porosity of 34%, exceeding the 31% recorded for the NiNCB-sim 800 °C. Moreover, NiNCNT-sim 800 °C catalyst layer exhibits a hierarchical pore network with a broad pore size distribution ranging from 2 to 278 nm, while the NiNCB-sim 800 °C is confined to a narrower pore size ranging from 2 to 154 nm (**Fig. 6d**).

The observed differences in 3D geometry and pore size translate directly to the simulated $CO_2$ molar-flux maps (**Fig. 6e, h**): NiNCNT-sim 800 °C catalyst layer shows an effective $CO_2$ diffusivity of $1.17\times10^{-5}$ m$^2$ s$^{-1}$. In contrast, the NiNCB-sim 800 °C electrode exhibits significantly lower diffusivity of $6.51\times10^{-6}$ m$^2$ s$^{-1}$, underscoring possible structural bottlenecks and mass-transport limitations. The reconstructed $CO_2$ velocity fields (**Fig. 6f, i**), obtained by solving the Navier-Stokes equation for steady-state, laminar, incompressible flow, reveal pronounced differences in the $CO_2$ transport behavior within the two catalyst layers. The NiNCNT-sim 800 °C catalyst layer exhibits continuous flow pathways, indicative of higher local permeability and reduced viscous resistance. On the other hand, the NiNCB-sim 800 °C layer leads to localized stagnation zones with severely restricted fluid penetration (**Fig. 6f**). Moreover, the gas permeability of the NiNCB-sim 800 °C is lower ($1.47\times10^{-16}$ m$^2$) than the NiNCNT-sim 800 °C catalyst layer ($3.87\times10^{-16}$ m$^2$). Higher $CO_2$ permeability shortens gas transport pathways and enables higher $CO_2$ concentration in the vicinity of active sites, contributing to improved CO selectivity as supported by $CO_2$R performance of the NiNCNT-sim 800 °C in MEA (**Fig. 1b**). Our simulation results also revealed that the effective electronic conductivity is significantly higher in the NiNCNT-sim 800 °C (669.29 S m$^{-1}$) compared to the NiNCB-sim 800 °C (148.00 S m$^{-1}$) catalyst layer (**Fig. 6g, j**).

To further understand how the architecture of the catalyst layer governs $CO_2$ transport and flooding/salt dynamics, we analyzed stability of the NiNCB-sim 800 °C electrode in the MEA under pulsed chronopotentiometry, alternating the current density between 200 mA cm$^{-2}$ for 3 s and 0 mA cm$^{-2}$ for 1 s. Cell voltage increased by 14% from the initial value of 2.89 V, while FE towards CO dropped to 78% from the starting value of 98 % within only 7 hours of electrolysis as illustrated in **Fig. S16**. On the other hand, the NiNCNT-sim 800 °C electrode maintained the performance for 70 hours, showing only a 10% change in cell voltage and a 11% decrease in FE towards CO (**Fig. S2a**). Our steady-state simulations results depict that the NiNCNT-sim 800 °C catalyst layer exhibits enhanced mass transport, continuous flow, and more uniform current distribution, highlighting the key role of larger interparticle porosity in the catalyst layer. In contrast, NiNCB-sim 800 °C catalyst layer suffers from structural bottlenecks, restricted $CO_2$ flux, and highly localized current densities, which accelerates degradation via salt precipitation, and flooding (48,49).

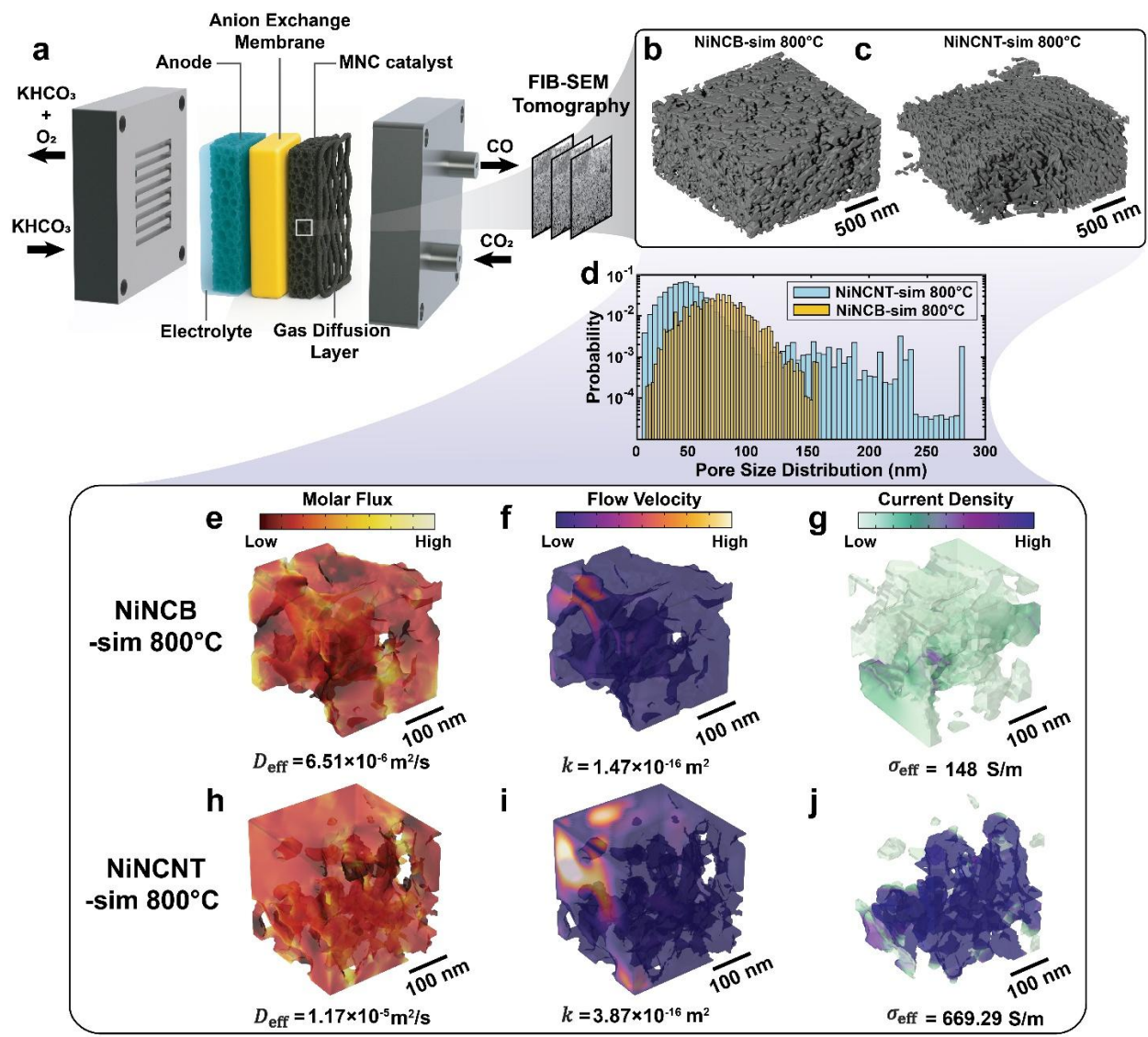

**Fig. 6 | a,** Schematic of the MEA electrolyzer system, 3D FIB-SEM reconstructions of $1 \times 2 \times 2$ μm$^3$ volumes for **b,** NiNCB-sim 800 °C, **c,** NiNCNT-sim 800°C catalyst layers, **d,** pore size histogram for , NiNCB-sim 800 °C and , NiNCNT-sim 800 °C based catalyst layers, COMSOL Multiphysics simulation results on $250 \times 250 \times 250$ nm$^3$ sub-volumes for **e,** CO$_2$ molar-flux distributions within the pores for NiNCB-sim 800 °C, **h,** NiNCNT-sim 800°C catalyst layers, **f,** simulation results for flow fields in NiNCB-sim 800 °C, **i,** NiNCNT-sim 800 °C samples, **g,** current density distribution results within the catalyst particles for NiNCB-sim 800 °C, **j,** NiNCNT-sim 800 °C catalyst layer. Color scales in all panels are normalized to their respective maximum.

**Catalyst cost estimation**

Despite the notion that MNC catalysts provide cost advantages over Ag catalysts for $CO_2R$ to CO, there does not exist a detailed cost analyses to provide quantitative justification for these claims. Given the complexity and uncertainty surrounding the commercial scalability of MNC catalysts, it is essential to evaluate their production costs at early stages. Similar to more established electrochemical technologies such as fuel cells, catalyst cost can represent a significant fraction of the total system cost in $CO_2$ electrolyzers, especially in low volume production scales (50). To this end we estimated the cost of NiNCNT-sim 800 °C developed in this work through the simultaneous wetness impregnation method (**Scheme 1**). More details on the cost estimation method can be found in the Supplementary Information (**Supplementary Note 2** and **Fig S17-20**). According to the CatCost (51) tool applied in this study, the projected cost of the NiNCNT-sim 800 °C catalyst based on the CapEx/OpEx method is substantially lower ($589.68 USD/kg) compared to that of the Ag-based cathode materials (~$1,900 USD/kg) reported in the literature for $CO_2R$ electrolyzers (52).

A Sankey diagram (**Fig. S21**) shows the NiNCNT-sim 800 °C projected purchase cost is dominated by the cost of carbon support and operating costs of the catalyst synthesis, highlighting a clear opportunity for cost reduction through using lower-cost carbon supports, and reducing the operating costs of the catalyst synthesis through process intensification. As an example, incorporating low-cost carbon supports with engineered pore structures, such as biomass-derived carbons (e.g., biochar) may offer viable pathways to reduce the catalyst cost (62). In parallel, engineering the support morphology to achieve higher interparticle porosity, as supported by our FIB-SEM/Multiphysics results, can enhance the $CO_2R$ performance. Sensitivity analysis also depicts the effect of different input parameters on the projected cost of producing NiNCNT-sim 800 °C catalyst in a dedicated factory (CapEx & OpEx method), as visualized through a tornado plot (**Fig. 7**). The high and low levels used for each parameter were estimated using literature resources and price surveys from the similar catalyst production units (53). Production scale, precursor cost (especially carbon support), and capital expenditure govern the overall catalyst cost, which aligns well with the cost distribution results visualized in the Sankey diagram (**Fig. S21**). This bottom-up cost analysis provides one of the first reported early-stage economic evaluations of an atomically dispersed MNC catalysts for electrochemical systems, offering critical insights into its cost drivers and scale-up viability compared to the precious metal-based alternatives. By quantifying material and processing costs, this work establishes a foundation for future cost optimization and system-level integration.

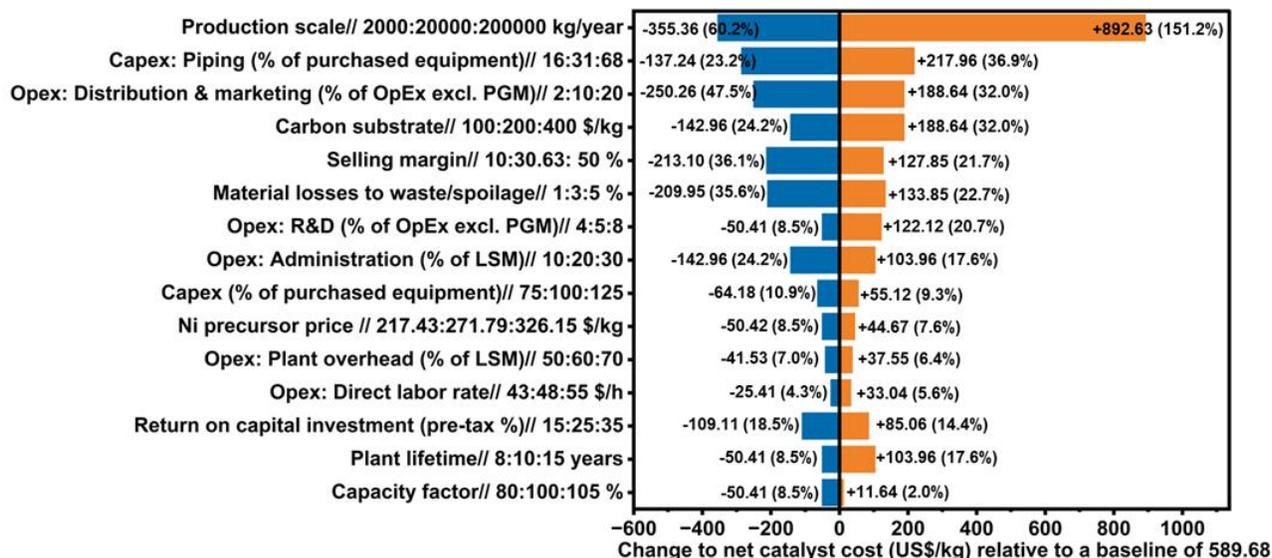

**Fig. 7** | Tornado plot showing the results of sensitivity analysis. The effect of key input parameters on the net catalyst cost for NiNCNT-sim 800 °C in a dedicated plant at an annual production scale of $2.0 \times 10^4$ kg. Input values are listed as low-cost input value: baseline input value: high-cost input value.

**Conclusions**

We uncovered three key factors governing the performance of NiNC catalyst synthesized in this work: the impregnation method of nitrogen precursor, the structure/morphology of the carbon support, and the chelating agent used during synthesis. Controlled impregnation of the nitrogen-precursor through simultaneous wetness impregnation method alongside the use of glucose as a chelating agent promote the uniform incorporation of N across the carbon structure and suppresses the agglomeration of Ni ions into clusters. By carefully tuning these parameters, the NiNCNT-sim 800 °C electrode achieved $j_{CO}$ of 558 mA cm$^{-2}$ at 3.2 V with 92% FE towards CO and $EE_{CO}$ of 39%, outperforming previously reported NiNC catalysts and Ag-based benchmarks under comparable MEA conditions. The coupled FIB-SEM tomography with numerical modeling results reveal facilitated $CO_2$ diffusion in the NiNCNT-sim 800 °C catalyst layer compared to NiNCB-sim 800 °C, which is associated to the larger interparticle porosity when using CNT as the catalyst support. Our results underscore that the mass transport in $CO_2R$ is not solely dictated by the intrinsic porosity of the carbon support but is also strongly correlated with the interparticle architecture within the catalyst layer. Achieving optimal $CO_2R$ performance requires an interplay between the nanoscale porosity of the catalyst support, which governs the concentration of electrochemically accessible active sites, and mesoscale interparticle spacing, which controls $CO_2$ diffusion and transport. Beyond $CO_2R$ electrolyzers, systems limited by gas transport and flooding include any catalyst layers supported on gas diffusion layers or porous transport layers, such as those used in fuel cells, metal-air batteries and water electrolyzers can benefit from the design metrics elucidated herein in terms of engineering the percolated meso/microporosity of the catalyst layer to reduce tortuosity to increase current densities and extend stability. We also estimate the projected purchase cost of the NiNCNT-sim 800 °C as $ 589 USD/kg, which is substantially lower than Ag-based catalysts, highlighting the industrial potential of this established synthesis method, as well as the demonstration of a quantitative route to identify potential cost advantages for emerging catalyst chemistries.

## Methods

### Materials and chemicals

Chemical precursors used for the catalyst synthesis, electrolyte and electrode preparation were purchased from Sigma-Aldrich. These include melamine, α-D-Glucose, high-purity potassium bicarbonate, Poly(tetrafluoroethylene) powder, nickel (II) nitrate hexahydrate, hydrochloric acid, isopropyl alcohol, and nitric acid. The multiwall CNTs were purchased from XFNANO (L:10-20 μm, OD: 4-6 nm). The CB (Ketjen Black) was purchased from Sigma-Aldrich. X37-50 Sustainion anion exchange membrane and iridium oxide anode were purchased from Dioxide Materials. Nafion ionomer D520 solution was also obtained from Ion Power, Inc. The gas diffusion layer (GDL, Sigracet 39 BB) was obtained from Fuel Cell Store. Milli-Q Type I ultrapure water (18.2MΩ·cm) was used for all experimental steps.

### Catalyst/ electrode synthesis

For synthesis of the NiNCNT/CB-phys catalysts, 0.15 g of carbon substrate (CNT or CB), 0.11 g of nickel (II) nitrate hexahydrate and 3 g of α-D-glucose were dispersed in 10 ml of Milli-Q Type I ultrapure water, and sonicated for 30 min to get a homogenous suspension, followed by stirring for 5 h at room temperature. The slurries were harvested after washing with water and drying at 70 °C. The resulting powder was ground and mixed with 0.75 g of melamine (mass ratio of 1:5 with carbon substrate), followed by pyrolysis at various temperatures (700-900 °C) under Ar atmosphere in a tube furnace for 2 hours using a heating rate of 5 °C/min. After pyrolysis, the sample was treated with a 5 wt.% HCl solution for 2 h (to remove possible Ni clusters), followed by thorough washing with Milli-Q Type I ultrapure water until the pH reached 6-7. The resulting material was then dried in an oven at 70 °C overnight. The same procedure was used for the synthesis of NiNCNT/CB-sim samples, with the key difference that all precursors (CNT, glucose, melamine, and the nickel precursor) were co-dispersed in 20 mL of Milli-Q Type I ultrapure water and stirred at 80 °C for 5 h to facilitate the dissolution of melamine. The electrodes were also fabricated by dispersing 4 mg of NiNC catalysts in a 2ml IPA solution containing 4 μL of Nafion ionomer and 0.52 mg of PTFE powder. This ink was drop-casted on the GDL to achieve a catalyst loading of 1 mg cm$^{-2}$, followed by drying in an oven for 4 hours at 60°C.

**Electrochemical tests**

All $CO_2R$ experiments were conducted in a 5 cm$^2$ MEA cell obtained from Dioxide Materials company with a serpentine channel. Components were compressed using eight bolts torqued to 40 lb·in. An anion exchange membrane (Sustainion X37-50 Grade RT, Dioxide Materials) was inserted between the cathode and the anode ($IrO_2$ coated on carbon paper). Throughout all experiments, 100 sccm of humidified $CO_2$ was fed into the cathode flow channels using a mass flow controller, while the anode side was fed with 0.5 M $KHCO_3$ (for long-cycle experiments, 0.1 M $KHCO_3$ solution was used as anolyte). Electrolyte was recirculated at 10 mL min$^{-1}$ with a peristaltic pump. Gaseous products were detected by online gas chromatography (SRI Multigas #5). To estimate the ECSA of the electrodes, the double-layer capacitance of the as-prepared electrodes was measured in Ar-purged 0.5 M $KHCO_3$ electrolyte. The scan rate was varied from 5 to 100 mV s$^{-1}$ in a non-Faradaic potential region using 3 mm glassy carbon electrode. Faradaic Efficiency (FE) for CO and $H_2$ was calculated based on the equation as follows:

$$FE_i = \frac{z_i \times x_i \times F}{Q} \times 100$$

Where $z_i$ is the number of electrons transferred for $CO_2R$ to CO/ or $H_2$, $x_i$ is the number of moles of products, F is Faraday's constant and Q is the total charge passed during the electrolysis. The full cell energy efficiency for $CO_2$ to CO in MEA was also calculated based on the following equation:

$$EE_{CO} = \frac{1.23 - (-0.10)}{E_{cell}} \times FE_{CO}$$

In this equation, $E_{cell}$ is the cell voltage applied between the cathode and anode in MEA without iR correction, $FE_{CO}$ is the measured Faradaic efficiency of CO in percentage, -0.10 V vs RHE is the equilibrium potential for $CO_2R$ to CO, and +1.23 V vs RHE accounts for OER equilibrium potential (54).

**Physicochemical characterization**

A Thermo Fisher Scientific Spectra Ultra (scanning) transmission electron microscope (STEM) with Cs probe, image correction and a high-brightness Schottky field emitter gun (X-FEG) with ultra-stable monochromator (UltiMono) source was used at 300 kV for transmission electron microscopy (TEM) analysis. A high-angle annular dark field (HAADF) detector and a Ceta camera were used for high-resolution STEM and TEM imaging, respectively. Energy dispersive X-ray (EDX) spectroscopy was done using the Ultra-X EDX system at a dwell time of 20 µs/pixel. Electron energy loss (EELS) was also acquired using CCD camera. Post-filtering of EDX maps were done by averaging every 3 pixels. Data analyses and curation were done using Velox (version 3.10.0.1130-5d766716c0), and Digital Micrograph Software. Specific surface area and pore structures of the catalysts were analyzed by nitrogen physisorption analysis using Quantachrome ASiQwin instrument. Before the measurements, degassing of the powder samples was done at 110 °C overnight under a continuous flow of $N_2$ gas. The surface area of the catalysts was estimated using a multipoint Brunauer-Emmett-Teller (BET) model.

X-ray photoelectron spectroscopy (XPS) measurements were performed using a Kratos AXIS Supra spectrometer equipped with a monochromatic Al Kα source (15 mA, 15 kV). All samples were analyzed using the Kratos charge neutralizer system. Survey spectra were acquired with a 300 µm × 700 µm analysis area and a pass energy of 160 eV, while high-resolution spectra were collected over the same area with a pass energy of 20 eV. ICP-OES (Agilent ICP-OES 7900) was used to determine the total Ni contents in the NiNC catalysts. X-ray diffraction (XRD) was performed using a Co Kα radiation source (λ = 1.79 Å) operating at 35 kV and 35 mA. X-ray absorption spectroscopy, comprising X-ray absorption near-edge spectroscopy (XANES) and extended X-ray absorption fine structure (EXAFS) at the Ni K-edge, was performed in total fluorescence yield mode using a silicon drift detector at beamline BL-32A of the Taiwan Photon Source (TPS), NSRRC. Energy calibration was performed using the first inflection point of the near-edge region of Ni foil. Raman spectroscopy was also performed using a Renishaw InVia Raman spectrometer equipped with a 500 mW HeNe laser ($\lambda$ = 532 nm), operated at 50 mW power with an 1800 L/mm grating.

**FIB-SEM tomography**

Focused ion beam-scanning electron microscopy (FIB-SEM) tomography was conducted on the NiNCNT-sim 800 °C and NiNCB-sim 800 °C cathodes using a ThermoFisher Helios 5 UC dual-beam microscope. Near-surface porosity in each unembedded sample was first filled using electron beam deposition of the platinum in the FIB at an accelerating voltage of 30 kV and current of 3.2 nA. This deposition helps reduce the depth of field to a single slice, while providing sufficient imaging contrast for the carbon-based matrix. Each site was then prepared using a standard means for FIB-SEM tomography (55). Tomography was conducted using Auto Slice-and-View software to obtain 1101 (NiNCB-sim 800 °C) and 1530 (NiNCNT-sim 800 °C) sequential images. Focused ion beam milling of each slice was conducted at 30 kV and 80 pA with an approximate slice thickness of 4 nm (NiNCB-sim 800 °C) or 2 nm (NiNCNT-sim 800 °C). Imaging of each slice was performed using secondary electron signal from a through-lens detector (TLD), with an electron beam accelerating voltage of 1.5-2 kV, a current of 50-1600 pA, a dwell time of 2-3 μs, twofold line averaging, and fourfold frame averaging. Post-process image stack alignment was done in Dragonfly 3D World using a Mutual Info slice alignment algorithm (1% initial translation, 0.01% final translation) and each image stack was then denoised using a 3D Gaussian blur (kernel size = 5, σ = 1.2). 2D UNet convolutional neural networks (CNNs) were trained to segment the images into infilled platinum deposition (as a porosity analog), catalyst material, and background classes using ten manually painted images as a ground truth (56). The batch size for training was 64, with a patch size of 48, a stride ratio of 0.80, and two-fold data augmentation. Trained CNNs were applied to the full dataset to segment the complete image stack into three classes. The segmented image stacks were then cropped to representative volumes with a size of 250 × 250 × 250 nm, and contour meshes were generated from the pore network segmentations using down-sampling to obtain ~ 10,000 surface mesh elements in each.

**Acknowledgements**

Z.T. acknowledges the financial supports from the NSERC Postdoctoral Fellowship. We acknowledge the support of the Government of Canada's New Frontiers in Research Fund (NFRF), CANSTOREnergy project NFRFT-2022-00197. We acknowledge the Canadian Centre for Electron Microscopy (CCEM) at McMaster University for access to instrumentation used to acquire HRTEM/STEM, and FIB-SEM tomography images. We also thank the staff of the McMaster Analytical X-Ray Diffraction Facility for assistance with XRD measurements, and Surface Science Western (SSW) center for XPS analysis. All XAS measurements were performed at beamline BL-32A of the Taiwan Photon Source (TPS). XRD measurements were conducted at McMaster Analytical X-ray Diffraction Facility (MAX). ICP-OES measurements were conducted at the Department of Materials Science and Engineering at McMaster University.